\documentclass[runningheads]{llncs}
\usepackage{hyperref}

\usepackage{multirow}
\usepackage{tabularx}
\usepackage{booktabs}

\usepackage{listings}
\lstdefinelanguage{JavaScript}{
  morekeywords={typeof, new, true, false, catch, function, return, null, catch, switch, var, if, in, while, do, else, case, break, let, for},
  morecomment=[s]{/*}{*/},
  morecomment=[l]//,
  morestring=[b]",
  morestring=[b]'
}
\usepackage{placeins} 

\usepackage{amsmath,amssymb}
\newtheorem{mydef}{Definition}
\newtheorem{mytheorem}{Theorem}

\usepackage{pifont} 

\usepackage{commath} 

\usepackage{soul} 
\usepackage{listings}
\usepackage{balance}
\usepackage{cleveref}
\crefname{mydef}{Definition}{Definitions}
\crefname{mytheorem}{Theorem}{Theorems}
\crefname{myeq}{Equation}{Equations}
\crefname{mylemma}{Lemma}{Lemmas}
\crefname{myclaim}{Claim}{Claims}
\crefname{myaxiom}{Axiom}{Axioms}
\crefname{myassumption}{Assumption}{Assumptions}
\crefname{myproposition}{Proposition}{Propositions}

\usepackage{arydshln} 
\usepackage{verbatim}

\usepackage{tikz}
\usepackage{forest}
\usepackage{subcaption}
\usepackage{svg}

\newcommand{\tradeoff}{trade-off}

\newcommand{\nonexpert}{non-expert}

\newcommand{\ep}{\varepsilon}
\newcommand{\eptext}{$\ep$}

\newcommand{\node}{node}

\newcommand{\respondentui}{{\textsc{Respondent UI}}}
\newcommand{\editor}{{\textsc{Poll Editor}}}
\newcommand{\resultui}{{\textsc{Server}}}
\newcommand{\datastruct}{\textit{Poll structure}}

\newcommand{\myquestion}{\textit{``How do you feel about your purchase?''}}
\newcommand{\questionanswers}{`\textit{Happy}', `\textit{Neutral}' and `\textit{Unhappy}'}
\newcommand{\followup}{\textit{``What's the reason you feel unhappy?''}}

\newcommand{\Randori}{{\textsc{Randori}}}
\newcommand{\airavat}{\textsc{Airavat}}
\newcommand{\gupt}{\textsc{Gupt}}
\newcommand{\opendp}{\textsc{OpenDP}}

\newcommand{\pythia}{\textsc{Pythia}}
\newcommand{\dpella}{\textsc{DPella}}
\newcommand{\pretpost}{\textsc{PreTPost}}
\newcommand{\pinq}{\textsc{Pinq}}
\newcommand{\fuzz}{\textsc{Fuzz}}
\newcommand{\rwpsi}{\textsc{Psi}}
\newcommand{\ektelo}{$\epsilon$\textsc{ktelo}}

\newcommand{\tool}{tool}
\newcommand{\tools}{tools}
\newcommand{\Tools}{Tools}

\usepackage{xcolor}
\colorlet{punct}{red!60!black}
\definecolor{background}{HTML}{EEEEEE}
\definecolor{delim}{RGB}{20,105,176}
\colorlet{numb}{magenta!60!black}

\lstdefinelanguage{json}{
    basicstyle=\normalfont\ttfamily,
    numbers=left,
    numberstyle=\scriptsize,
    stepnumber=1,
    numbersep=8pt,
    showstringspaces=false,
    breaklines=true,
    frame=lines,
    backgroundcolor=\color{background},
    literate=
     *{0}{{{\color{numb}0}}}{1}
      {1}{{{\color{numb}1}}}{1}
      {2}{{{\color{numb}2}}}{1}
      {3}{{{\color{numb}3}}}{1}
      {4}{{{\color{numb}4}}}{1}
      {5}{{{\color{numb}5}}}{1}
      {6}{{{\color{numb}6}}}{1}
      {7}{{{\color{numb}7}}}{1}
      {8}{{{\color{numb}8}}}{1}
      {9}{{{\color{numb}9}}}{1}
      {:}{{{\color{punct}{:}}}}{1}
      {,}{{{\color{punct}{,}}}}{1}
      {\{}{{{\color{delim}{\{}}}}{1}
      {\}}{{{\color{delim}{\}}}}}{1}
      {[}{{{\color{delim}{[}}}}{1}
      {]}{{{\color{delim}{]}}}}{1},
}

\begin{document}
\title{Randori: Local Differential Privacy for All}
\titlerunning{Randori}
%
\author{Boel Nelson}
\authorrunning{B. Nelson}
%
\institute{Chalmers University of Technology, Gothenburg 412 96, Sweden\\
\email{boeln@chalmers.se}}
\maketitle              
\begin{abstract}
Polls are a common way of collecting data, including product reviews and feedback forms. However, few data collectors give upfront privacy guarantees. Additionally, when privacy guarantees are given upfront, they are often vague claims about `anonymity'. Instead, we propose giving quantifiable privacy guarantees through the statistical notion of \textit{differential privacy}. Nevertheless, privacy does not come for free. At the heart of differential privacy lies an inherent \tradeoff\ between accuracy and privacy that needs to be balanced. Thus, it is vital to properly adjust the accuracy-privacy \tradeoff\ before setting out to collect data.

\hspace{1em}
Motivated by the lack of \tools\ to gather poll data under differential privacy, we set out to engineer our own \tool. Specifically, to make \textit{local differential privacy} accessible for all, in this systems paper we present \Randori, a set of novel open source \tools\ for differentially private poll data collection. \Randori\ is intended to help data analysts keep their focus on \textit{what} data their poll is collecting, as opposed to \textit{how} they should collect it. Our \tools\ also allow the data analysts to analytically predict the accuracy of their poll. Furthermore, we show that differential privacy alone is not enough to achieve end-to-end privacy in a server-client setting. Consequently, we also investigate and mitigate implicit data leaks in \Randori.
\end{abstract}
\section{Introduction}\label{sec:introduction}
Polls are a widely used way of collecting data. For example, one might be asked to fill out a review after purchasing an item online. Now, these polls can consist of an arbitrary number of intertwined questions. For example, after purchasing an item online, one might be asked \myquestion\ with answer alternatives \questionanswers. Now, the merchant can also choose to add a follow-up question asking \followup\ to all respondents that answer that they were unhappy with their purchase. That is, polls can become complex depending on the number of questions and follow-up questions.

Having established that polls are indeed an interesting way to gather data, we ask ourselves how we could gather such data while also providing our respondents with useful privacy guarantees. A popular topic within the area of privacy-preserving data collection is \textit{differential privacy}. Differential privacy is a rigorous statistical notion of privacy where privacy loss is quantified. To achieve privacy data is perturbed, usually through injecting data with controlled noise. As such, at the core of differential privacy lies an inherent trade-off between accuracy and privacy.

While differential privacy is widely accepted as a strong notion of privacy, it is mainly used in prototypes created by academics. So far, differential privacy in real, deployed systems, is only observed at tech giants such as Apple~\cite{thakurta_learning_2017-1,thakurta_emoji_2017}, Google~\cite{erlingsson_rappor_2014} and Microsoft~\cite{ding_collecting_2017}.
The US Census Bureau~\cite{garfinkel_issues_2018-1} also point out that there are several issues such as setting the privacy parameter (\eptext) and the lack of tools to verify the correctness of the implementation of differential privacy. We want to provide a tool that tackles the privacy parameter from an accuracy perspective, and that at the same time guarantees differential privacy. As such, we want to offer differential privacy for all interested in collecting poll data.

In order to successfully apply differential privacy to data collection, an analyst must not only correctly implement the noise injection, but also balance the accuracy-privacy trade-off for their particular data. What's more, allowing arbitrarily intertwined data through follow-up questions creates additional challenges. From the example before, having answered the follow-up question leaks that the respondent felt unhappy about their purchase. Put differently, the poll structure has the ability to create unintentional, implicit information flows. As such, the data collection process can also contain \textit{side-channels} that leak information, all of which are not captured by differential privacy itself.

Consequently, we identify three main problems with collecting poll data under differential privacy:

\begin{itemize}
    \item Implementation needs to be correct
    \item Gathered data may be too inaccurate
    \item Side-channels may arise during collection
\end{itemize}

To make differential privacy accessible for all, we present a novel set of open source \tools\ called \Randori~\footnote{Source code available at (anonymous account):\\ \url{https://www.dropbox.com/sh/kb0tvtt17eszdvb/AACC8RLsLjQlxUh14\_\_aHK1ea}}. \Randori\ helps a data analyst first design a poll, tune the accuracy-privacy trade-off, and later collect data from respondents under differential privacy. What's more, we have investigated and addressed side-channels that can arise when a respondent receives and answers a poll. These side-channels include, but are not limited to, implicit leaks from follow-up questions. As such, not only are we interested in accurate differentially private data collection, but we also protect privacy end-to-end throughout the entire collection process. By presenting \Randori\, our contributions are:

\begin{itemize}
    \item[\textbf{+}] \Tools\ for \textit{designing polls}, and \textit{collecting data} under differential privacy
    \item[\textbf{+}] A \tool\ for \textit{predicting} and \textit{tuning accuracy} for a given poll
    \item[\textbf{+}] A data collection process that is \textit{end-to-end private}
\end{itemize}

\section{Differential Privacy}\label{sec:dp}
Differential privacy is a statistical notion of privacy that represents a property of an algorithm, as opposed to being a property of data. As such, differential privacy is fundamentally different from privacy models such as k-anonymity~\cite{samarati_protecting_1998}, l-diversity~\cite{machanavajjhala_l-diversity_2007} and t-closeness~\cite{li_t-closeness_2007}, where privacy guarantees are derived from what values are present in the data.

In this paper we focus specifically on \textit{local differential privacy}~\cite{kasiviswanathan_what_2011}, which is relevant whenever the mechanism is applied locally at the data source rather than centrally. In the rest of the paper, when we talk about differential privacy, we mean specifically \textit{local} differential privacy.
\begin{mydef}[$\ep$-Differential Privacy]\label{def:eDP}
A randomized algorithm $\mathcal{M}$, with an input domain $\mathcal A$ and an output domain $\mathcal{B}$, is
$\ep$-differentially private if for all possible inputs $a$ and $a'$, and all possible output values $b$,
\[
\text{Pr}[\mathcal{M}(a) = b]\leq e^{\ep} \times \text{Pr}[\mathcal{M}(a') =b].
\]
\end{mydef}

\vspace{-0.5em}



The core mechanism used in this paper to achieve differential privacy is a variant of the classic \textit{randomized response} algorithm~\cite{warner_randomized_1965}. Using a binary input domain (`yes' or `no'), the randomized response algorithm can be described as follows: flip a coin $t$. If $t$ lands heads up then respond with the true answer (the input). Otherwise flip a second coin $r$ and return `yes' if heads, and `no' if tails. 

Basically, this algorithm will either deliver the true answer, or randomly choose one of the viable answers. By delivering an answer in this way, we say that the respondent enjoys \textit{plausible deniability}. That is, a respondent may always claim that their response was different from their true answer, without anyone being able to disprove their claim.

In this algorithm the bias of the coin $t$ determines the privacy-accuracy \tradeoff, whereas the coin $r$ can always be unbiased (i.e. it has a uniform distribution). The variant of this mechanism used in this paper is a simple generalization: it (i) allows for a non binary input domain, and (ii) permits the bias of the coin $t$ to be dependent on the value of the input.   

\begin{mydef}[Randomized Response]\label{def:rr}
Let $A$ be the data domain, and $T = \{t_a\}_{a\in A}$ be an indexed set of values in $[0,1]$. Given these, we define the \emph{randomized response mechanism} $\mathcal M_T$, a randomized function from $A$ to $A$, as follows:
\newcommand{\Prob}[1]{\text{Pr}[#1]}
\[
\Prob{\mathcal M_T(a) = b} = \begin{cases}
    t_a + r_a & \text{when $a = b$},
\\  r_a       & \text{otherwise}
\end{cases}
\] where $r_a = (1 - t_a)/|A|.$
 \end{mydef}

A given randomized response mechanism $M_T$ can be viewed as a transition matrix. For example, suppose we have $A = \{a_1,a_2,a_3\}$ and $T = \{t_1,t_2,t_3\}$, then $M_T$ can be represented as the transition matrix:
\begin{align*}
M_T = &\bordermatrix{~ &a_1 & a_2 & a_3 \cr
                  a_1 & t_1+r_1 & r_1 & r_1 \cr
                  a_2 & r_2 & t_2+r_2 & r_2 \cr
                  a_3 &r_3 & r_3 & t_3+r_3 \cr}
\end{align*}
Note that each row sums to one (by definition of the $r_i$).
The definition of differential privacy essentially compares any two values drawn from the same column of the matrix, so \eptext\ is determined by the worst case among these:
\begin{mytheorem}[Differential Privacy of RR]
\label{thm:DPRR}
Randomized response mechanism $\mathcal M_T$ is
\eptext-differentially private, where $e^\ep$ 
is given by maximizing the value of $(t_a+r_a)/r_b$ over $a,b \in A$ where $a \neq b$. 
\end{mytheorem}


At this point it is appropriate to mention why we allow the `truth' probability $t_a$ to be different for each input: we want this additional degree of control because we will support mechanisms built from making follow-up questions, as in ``Do you smoke?'' then if the answer is yes, follow with ``More than 10 per day?''. Since the follow-up question will effectively increase the precision of the `yes' answer, we may choose to lower the initial precision of the `yes' answer by lowering the value of $t_{\text{`yes'}}$. We say that we allow for \textit{non-uniform diagonals} in our transition matrix. Later, in \Cref{sec:discussion}, we elaborate on the challenges non-uniformity creates when predicting accuracy.

Now, we also want to be able to reason about the accuracy of our algorithm. Deriving from the concept of a misclassification rate~\cite{kasiviswanathan_what_2011}, we define our metric for error in \Cref{def:error}. That is, if a response $a$ gets mapped by the randomized response mechanism to a value other than itself, it is considered misclassified.
\begin{mydef}
[Error Metric]\label{def:error}
Let $\mathcal{M_T}$ represent randomized response, then given for any answer $a \in \mathcal{A}$ the error is the probability of outputting any other output in $\mathcal{A}$:
\[X_a = \text{Pr}[\mathcal{M_T}(a) \neq a] \]
\end{mydef}
A general analytical error bound~\cite{dwork_differential_2008}, for any algorithm, is given by the Chernoff bound in \Cref{def:accuracy}. We say that an algorithm is ($\alpha, \beta$)-useful~\cite{zhu_differential_2017}. 

\begin{mytheorem}[Analytical Accuracy]\label{def:accuracy}
Let $X$ be a random variable representing the error of the output of a differentially private algorithm, and $\alpha$, $\beta$ two error bounds, then with probability 1-$\beta$, the error $X$ is bounded by at most error $\alpha$: 
\[ \text{Pr}[X \leq \alpha] \geq 1-\beta\]
\end{mytheorem}
\vspace{-0.5em}

Next, we express $\beta$ in terms of $\alpha$ and include the population size $n$ in the equation by using additive Chernoff bounds:
\vspace{-0.5em}
\begin{align*}
    \text{Pr}[X \leq \alpha] \geq 1 - 2e^{-2 \alpha^2 n}
\end{align*}
\vspace{-0.5em}

\section{Methodology and Assumptions}\label{sec:method}
In order to show that it is possible to build a \tool\ for differentially private data collection that is end-to-end private in a server-client setting, we construct a proof of concept called \Randori. Our intended goal is to build a prototype that works, not necessarily build one that is optimal from an accuracy, usability or performance aspect. Hence, our main focus is to make sure \Randori's data collection is indeed differentially private, and that the process itself is protected against information leakage. To limit our possible implementations, we have constructed a list of functional requirements for \Randori\ in \Cref{sec:reqs}.

Recalling our example with a customer scoring a product they purchased online, \Randori\ must protect all of the customers' true answers. Note that differential privacy guarantees that any two outputs are ($\epsilon$-) indistinguishable, but since we capture and process the data in an interactive setting, we also need to make sure there are no implicit leaks. That is, we will also examine what data could potentially leak from the server-client setting. In order to make a thorough investigation, we introduce a threat model in \Cref{sec:model}.


\subsection{Functionality Requirements}\label{sec:reqs}
Our goal is to make a system that is I) differentially private by design, II) able to predict error and III) protected against side-channels. We break these goals down into requirements as follows: 
\vspace{-.5em}
\begin{enumerate}
    \item \textit{Randomized response should be implemented automatically. Specifically, implementation details should be hidden from the data analyst.} \label{req:rr}
    \item \eptext\ \textit{should be calculated automatically:} \label{req:epsilon} \begin{enumerate}
        \item \textit{For the data analyst}
        \item \textit{For the respondent, since they do not trust the data analyst}
    \end{enumerate}
    \item \textit{The data analyst should be able to predict the error of a poll} \label{req:error}
    \item T\textit{he data analyst should be able to tweak the predicted error by changing the probability ($t$) of randomly choosing each answer} \label{req:probability}
    \item \textit{Statistical noise should be filtered away automatically} \label{req:noise}
    \item \textit{Several data analysts should be able to cooperate, eliminating the need for every analyst to be a privacy expert themselves} \label{req:users}
\end{enumerate}
\vspace{-.5em}
Essentially, we want to remove most of the tedious details of implementing a differentially private algorithm, and let the data analysts focus on \textit{what} data to collect, instead of \textit{how} to collect it. With \Cref{req:rr} and \Cref{req:noise}, we hide the implementation of randomized response under the hood of \Randori.

Still, the value of \eptext\ is something we cannot abstract away from the system completely. That is, we cannot set \eptext\ to a `good' value since `good' is subjective and highly domain dependent. In the same way, error tolerance is also distinctly domain dependent. For example, $\pm$ 10g might be an acceptable error range when constructing scales for people, but not for scales used in medicine manufacturing. Hence, instead of trying to abstract away \eptext, we let the data analyst tune the privacy/accuracy trade-off themselves. The privacy/accuracy tuning is achieved through a combination of \Cref{req:epsilon}, \Cref{req:error} and \Cref{req:probability}.

Furthermore, we argue that \Cref{req:users} is important to make \Randori\ more accessible in our case. Namely, by allowing multiple users, not every data analyst needs to be a privacy expert. That is, any data analyst can design the poll and collect the data without breaking differential privacy. Still, at least one user should be a privacy expert in the sense that someone needs to approve of the privacy/accuracy balance achieved.
\subsection{Threat Model and System Privacy Guarantees}\label{sec:model}

\textbf{Adversary Model and Assumptions.} We assume that adversaries can be either passive or active. The active adversary can send out polls using \Randori. Consequently, we assume that the adversary can pose as a data analyst. 
The passive adversary can observe and read the contents of all network traffic between data analyst and respondent. That is, we consider both the case where the communication takes place in plain text, and the case where the adversary is strong enough to break any encryption used during communication. That is, we assume an adversary that can read message contents even when the communication is done over HTTPS. Still, we assume that the internal state of the code the respondent is running and the respondent's hardware cannot be monitored by the adversary. That is, the respondent is entering their true answers into a trusted computing base.

We also assume that the respondent does not close their client before our code has finished executing. Later, we elaborate on ways to handle non-termination and the challenges of implementing these defenses in our discussion (\Cref{sec:discussion}).

We do not consider cases where the respondent is an adversary that tries to attack the accuracy of the poll by skewing their answers. That is, we will only consider attacks on privacy, and not attacks on accuracy. 

\textbf{Trust.}
The sensitive data in this setting is the respondents' \textit{true answers} to polls. That is, \textit{responses} produced by randomized response are not considered sensitive as the respondent enjoys \textit{plausible deniability}. Hence, sensitive data only resides in the respondent's application.

Moreover, we consider the code running on the respondent's device to be completely trusted by the respondent. That is, the code the respondent is running is allowed to hold and take decisions based on sensitive data.

As for the data analysts, we do not consider any of their data to be sensitive. Consequently, the poll questions are considered public data. Hence, the \eptext\ for any poll is also public data. We will also assume that the value of each respondent's privacy budget is public. That is, whether or not a respondent has participated in a poll also becomes public. We do not attempt to hide the identity of the respondents, but settle for plausible deniability.

Furthermore, the data analysts are considered untrusted by the respondent. That is, the respondent only wants to share their poll answers under differential privacy, and do not wish to share any other data than what is publicly known with the data analysts.



\textbf{System Guarantees.} We guarantee that the respondent will enjoy differential privacy, meaning we assure plausible deniability. If the data analyst turns out to be an adversary, the respondent still has plausible deniability, which we will not consider a privacy breach. We will guarantee that no matter the parameters fed to \Randori, differential privacy still holds. For example, the probability of answering truthfully is guarded by a \textit{truth threshold value}, since answering 100\% truthfully provides no privacy. However, we do not guarantee that the value of the truth threshold nor the value of the respondent's privacy budget are set to \textit{useful values}: we merely enforce them.

As for randomness, we strive to make a best effort given the programming language used. Hence, our implementation of differential privacy uses Javascript's Crypto library to ensure cryptographically strong random values~\cite{mozilla_and_individual_contributors_cryptogetrandomvalues_2020}, which is the strongest implementation of randomness available in JavaScript.

Furthermore, we guarantee that \Randori\ protects against a number of side-channels that would break end-to-end privacy. To the best of our knowledge, these are all side-channels under our given threat model.

\textbf{System Limitations.} In \Randori, the respondents do not have a persistent application. Hence, we cannot store the privacy budget between sessions. Instead, we assume a maximum budget \textit{per poll}. 

In its current state, \Randori\ does not contain a trusted third party to send the \respondentui\ to the respondents. Still, adding a third party only requires a minor change where the respondent visits the trusted third party to receive the \respondentui, but polls can still be served by the untrusted data analyst.

We do not consider the identity of the respondents to be secret, and thus we do not protect against leaking information through participation alone.

Also, we do not guarantee security through encryption, since we assume an adversary strong enough to break encryption. Still, we expect the users to encrypt the communication and take adequate measures to store the collected data securely, but we leave this outside of our system scope.
\vspace{-1.5em}
\section{Randori}\label{sec:randori}\vspace{-0.5em}
\Randori\ is a set of \tools\ with two focal points as far as functionality goes. These focal points are: \textit{poll design} and \textit{data collection}. In this section we will both describe the functionality of \Randori, the \tools\ it consists of, as well as how differential privacy is achieved. Lastly, we describe the steps taken to assure end-to-end privacy, as this property is not captured by differential privacy itself.
\vspace{-1.5em}
\subsection{\Tools\ and Vision}
\begin{figure}[htb]
    \centering
    \includegraphics[scale=0.12]{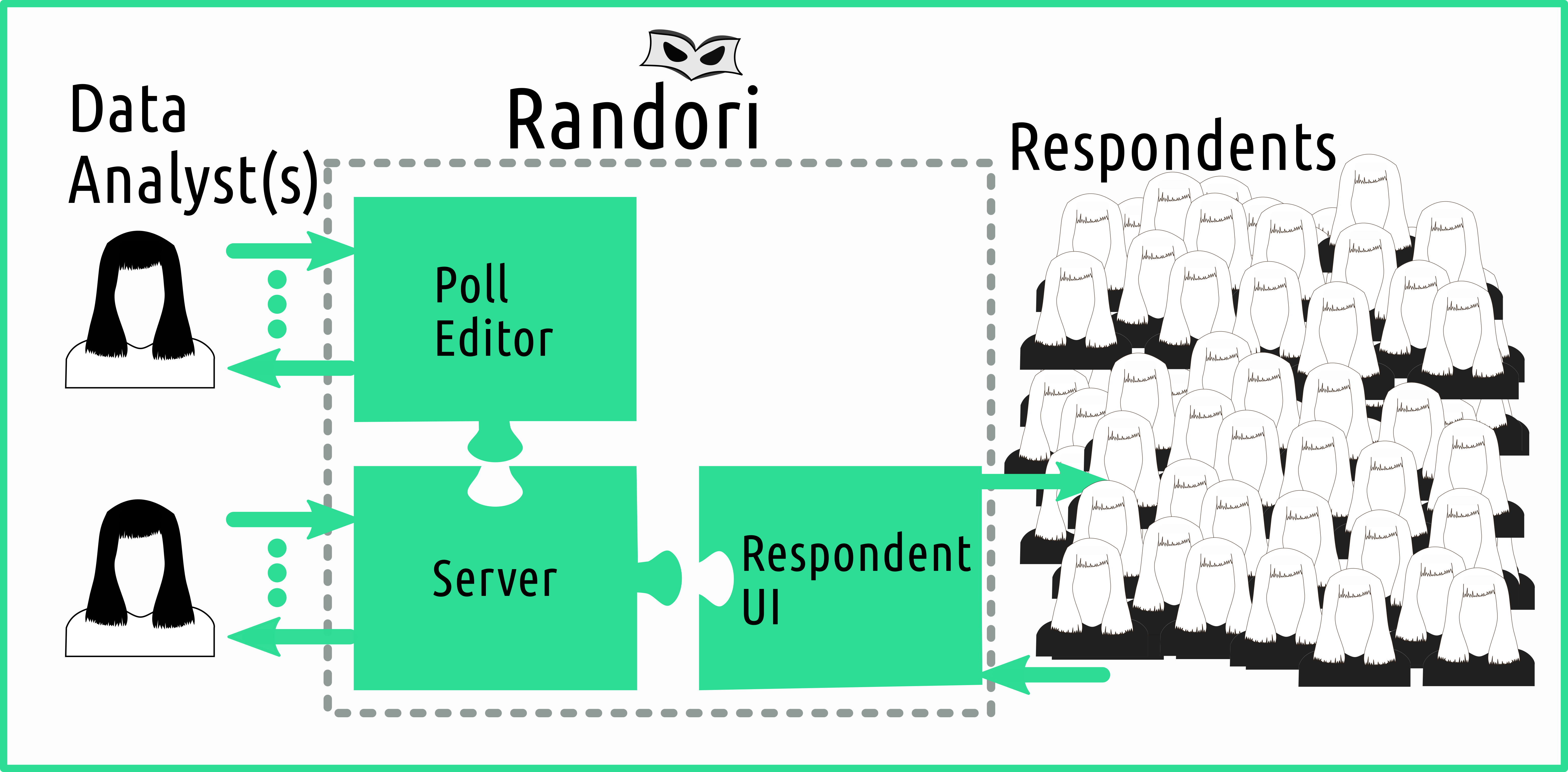}
    \caption{The different \tools\ included in \Randori\vspace{-2.5em}}
    \label{fig:concept}
\end{figure}

\Randori\ is a set of \tools\ (\Cref{fig:concept}) that enable data analysts to first design, and then perform data collections under differential privacy. 

We allow for multiple data analysts (Requirement \cref{req:users}) by first focusing the scope of each \tool, and secondly by choosing a portable data format that can easily be imported and exported by each \tool.
\vspace{-0.5em}
\subsubsection{Poll Design: \editor}
The \editor\ is where the data analyst creates and edits the poll structure and content. To be able to hide details under the hood it consists of two modes: \textit{edit} and \textit{explore}. 

\vspace{-1.5em}

In the \textit{edit mode} (screenshot in Appendix \Cref{fig:edit}) the data analyst can focus solely on the poll content: order of questions, number of answers and what answers trigger follow-up questions. That is, in the \textit{edit mode} it is as if the data analysts is editing any poll, not a special poll that is to be gathered under differential privacy. Polls are imported/exported on our JSON format (Appendix \Cref{fig:json}). 

Then, the fact that data is to be gathered under differential privacy is visible in the \textit{explore mode} (screenshot in Appendix \Cref{fig:explore}).  Arguably, without adequate accuracy, collected data becomes useless to the data analyst. To mitigate the problem of high error, we let the data analysts explore the accuracy-privacy trade-off through two sets of parameters: (i) \textit{True/Random answers} and  \textit{Weight},  and (ii) \textit{Alpha}, \textit{Beta} and \textit{Population}.

The parameters from set (i) influence the value of \eptext, where the slider is a course-grained adjustment affecting \textit{all} answers ($t_1, ..., t_n$ from transition matrix), and weight is a fine-grained adjustment available per answer (affecting a specific $t_a$). However, the analysts never directly sets any of the values in the transition matrix. Instead, $t_a$ is calculated from the product of the truth/random ratio, every parent's weight and the answer's own weight.
 
All parameters from set (ii) are part of the Chernoff bound and are calculated using \Cref{eq:equation-system}. The parameters \textit{population}, \textit{alpha} and \textit{beta} are shown on a per answer basis. The data analyst is required to set values for two of the parameters per answer, and the \editor\ calculates the third parameter. For example, if the analyst requires a certain accuracy (\textit{alpha}), they can set \textit{alpha} and explore what values of \textit{population} and \textit{beta} gives them their desired accuracy. Hence, the \editor\ implements Requirements \cref{req:error} and \cref{req:probability}.

Based on Vadhan~\cite{vadhan_complexity_2017} we construct the following equation system to display the relationship between $\alpha$, $\beta$, $n$ and \eptext.
\begin{equation*}\label{eq:equation-system}
\begin{cases}
    \alpha =  \frac{1+e^\epsilon}{e^\epsilon-1} \lambda\\
    \beta = {2e}^{-2{\lambda^2}n}\\
    \ep = \log(\frac{\frac{-\alpha}{\lambda}-1}{1-\frac{\alpha}{\lambda}})\\
    n = \frac{(1 + e^\ep)^2 \log(2/\beta)}{2\alpha^2 (e^\ep - 1)^2}\\
\end{cases}
\text{where }\lambda = \sqrt{\frac{\log{\frac{2}{\beta}}}{2n}}
\end{equation*}
\vspace{-2.5em}

\subsubsection{Data Collection: \resultui}
The \resultui\ holds the currently active poll on our JSON format. The \respondentui\ then accesses the poll from e.g. \texttt{localhost: 5000/poll}. Next, data analysts can access poll results through e.g \texttt{localhost: 5000/results}. The server post-processes responses from the respondents by filtering away statistical noise using Bayes' theorem. The results are shown to the data analysts in form of a JSON file. Hence, we fulfill Requirement \cref{req:noise}.
\vspace{-0.5em}
\subsubsection{Data Collection: \respondentui}
The \respondentui\ (screenshot in Appendix \Cref{fig:respondent-ui}) is a JavaScript client running on the respondents' device. As the \respondentui\ is trusted by the respondent, it can branch on sensitive data to present the respondent with questions based on their previous answers. Hence, to the respondent, a \Randori\ polls looks just like any other poll, but randomized response runs in the background. Hence, the \respondentui\ also takes care of Requirement \cref{req:rr}. Also, since the respondents do not trust entities outside of the \respondentui, the \respondentui\ re-calculates \eptext. 
\vspace{-0.5em}
\subsection{Differential Privacy}
Differential privacy is achieved in \Randori\ through randomized response. Since we are in the local setting, ensuring differential privacy is entirely done by the \respondentui. In particular, we ensure differential privacy through the two following practices:\\
\vspace{-0.5em}
\begin{itemize}
    \item Use of a strong random number generator client-side (\Cref{sec:rr})
    \item Data representation that prevents information leakage from follow-up questions (\Cref{sec:leakage})
\end{itemize}
\vspace{-0.5em}
\subsubsection{Implementation of Randomized Response}\label{sec:rr}
The strongest implementation of randomness in JavaScript is a cryptographically random number generator. Hence, we will use JavaScript's crypto library in our implementation. 
We show our implementation mainly as pseudo-code in \Cref{code:rr} in the Appendix.

Then, we calculate the corresponding value of \eptext\ (\Cref{code:epsilon}). Here, we find the biggest possible ratio, \textit{when changing any answer}, between the probabilities in the transition matrix. That is, we find the biggest possible ratio \textit{per column}.

Before we let the respondent answer the poll, we check that they have enough budget left for the full poll. We have also introduced a \textit{truth threshold} (here set to a dummy value of 0.99), as even polls with 100\% of truthful answers would otherwise be considered valid polls. The corresponding validity checks are shown in \Cref{code:budget}.

\subsubsection{Mitigating Structural Information Leakage}\label{sec:leakage}
Next, we address implicit information leaks. In particular, we investigate how we can allow for follow-up questions without letting them leak information about their parent questions. Recalling our example from the introduction with the question \myquestion, only respondents that answer `\textit{Unhappy}' get a follow-up question. Accordingly, \textit{any answer to the follow-up question} leaks that the first answer was `\textit{Unhappy}'. We want to ensure that \Randori\ can handle follow-up questions without leaking answers to other questions.

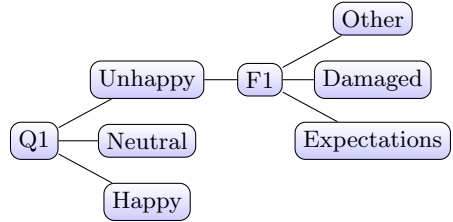
\begin{figure}[htb]
    \centering
\begin{tikzpicture}[sibling distance=2.5em, grow=0,
  every node/.style = {rounded corners,
    draw, align=center,
    top color=white, bottom color=blue!20}]]
  \node {Q1}
    child { node {Happy} }
    child { node {Neutral} }
    child { node {Unhappy}
      child { [sibling distance=2.5em] node {F1}
        child { node {Expectations} }
        child { node {Damaged} }
        child { node {Other} } }};
\end{tikzpicture}
\caption{\textbf{Version 1: }Naive representation of follow-up questions, any answer to \followup\ (F1) reveals that the respondent answered `\textit{Unhappy}'}
    \label{fig:naive-follow-up}
\end{figure}

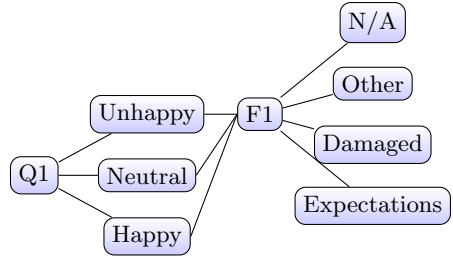
\begin{figure}[htb]
\centering
\begin{tikzpicture}[sibling distance=2.5em,grow=0,
  every node/.style = {rounded corners,
    draw, align=center,
    top color=white, bottom color=blue!20}]]
  \node {Q1}
    child { node (H) {Happy} }
    child { node (N) {Neutral} }
    child { node {Unhappy}
      child { [sibling distance=2.5em] node (F1) {F1}
        child { node {Expectations} }
        child { node {Damaged} }
        child { node {Other} } 
        child { node {N/A} } }};
    \draw (H.east) -- (F1.west);
    \draw (N.east) -- (F1.west);
\end{tikzpicture}
    \caption{\textbf{Version 2: }Follow-up with no information leaks, i.e. the F1 has valid answers for all respondents}
    \label{fig:follow-up}
\end{figure}

\begin{figure}[htb]
    \centering
\begin{tikzpicture}[sibling distance=2.5em, grow=0,
  every node/.style = {rounded corners,
    draw, align=center,
    top color=white, bottom color=blue!20}]]
  \node {Q1}
    child { node (H) {Happy} }
    child { node (N) {Neutral} }
    child { node [dashed] {Unhappy}
      child { [sibling distance=3.5em] node (F1) {F1}
        child { node {Unhappy\\Expectations} }
        child { node {Unhappy\\Damaged} }
        child { node {Unhappy\\Other} } }};
\end{tikzpicture}
    \caption{\textbf{Version 3: }Follow-up with less answers than previous implementation to preserve privacy budget}
    \label{fig:follow-up-improved}
\end{figure}
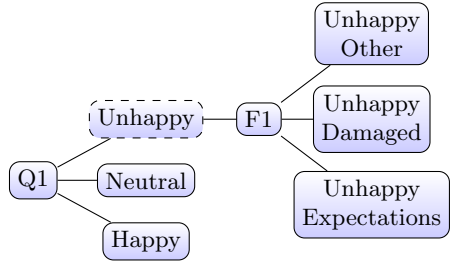

\begin{enumerate}
    \item[$\triangleright$] \textbf{Implementation:} \datastruct\ Version 1 (\Cref{fig:naive-follow-up})
    \item[\ding{55}] \textbf{Problem:} By allowing first \questionanswers\ to be sent, then any reply reveals that the respondent answered that they were unhappy with their purchase. We need to make sure that the number of replies sent is indistinguishable no matter the actual answers.
\end{enumerate}
\vspace{-1.75em}
\vspace{-1.75em}
\begin{enumerate}
    \item[$\triangleright$] \textbf{Implementation:} \datastruct\ Version 2 (\Cref{fig:follow-up})
    \item[$\triangleright$] \textbf{Reasoning:} We introduce an answer that is always valid for respondent's that did not trigger the follow-up question, namely `\textit{N/A}'.
    \item[\ding{55}] \textbf{Problem:} The number of replies is indistinguishable, but we spend budget on `\textit{N/A}' which in this case implies \textit{not 'Unhappy'}. Note that we cannot determine if `\textit{N/A}' means '\textit{Happy}' or '\textit{Neutral}' in this case, so we are not able to achieve any additional accuracy even though we spent privacy budget on `\textit{N/A}'. As such, while we do not break differential privacy, we end up wasting our budget by having to add `\textit{N/A}' \textit{for every} follow-up question.
\end{enumerate}
\vspace{-1.75em}
\vspace{-1.75em}
\begin{enumerate}
    \item[$\triangleright$] \textbf{Implementation:} \datastruct\ Version 3 (\Cref{fig:follow-up-improved})
    \item[$\triangleright$] \textbf{Reasoning:} In this case, we represent the tree logically in the same way as version 2, but we only send one response. That is, the dotted nodes represent answers that the respondent can choose, but that are never sent. Basically, we allow for the respondent to traverse the tree and choose an answer to each triggered question, but in reality the client will only send one response to each subtree (a question and all its follow-ups) of the poll. Since the client is a trusted component, branching on secrets is ok, so we can do this without leaking information. In this example, the respondents that choose `\textit{Happy}' or `\textit{Neutral}', are never asked why they were unhappy (which they were not), but the server never learns this due to the use of a single response.
\end{enumerate}

\vspace{-1.75em}

\FloatBarrier

\subsection{End-to-End Privacy}
Simply implementing randomized response to deliver responses is not enough to protect respondents' privacy, since the data collection process is may leak additional information. For example, information leakage through side-channels such as differences in timing, is not captured by randomized response. Consequently, to achieve end-to-end privacy, we need to make sure \Randori\ protects against side-channels. Next, we iterate through our implementation of the \respondentui\ to show the side-channels we have encountered.

\begin{figure}[htb]
     \centering
     \begin{subfigure}[t]{0.47\textwidth}
         \centering
         \input{code/v1}
     \end{subfigure}
     \hfill
     \begin{subfigure}[t]{0.5\textwidth}
         \centering
         \input{code/v2}
     \end{subfigure}
\end{figure}

\begin{figure}[htb]
     \centering
     \begin{subfigure}[t]{0.5\textwidth}
         \centering
         \input{code/v3}
     \end{subfigure}
     \hfill
     \begin{subfigure}[t]{0.47\textwidth}
         \centering
         \input{code/v4}
     \end{subfigure}
\end{figure}
\begin{enumerate}
    \item[$\triangleright$] \textbf{Implementation:} Version 1 (\Cref{code:v1})
    \item[\ding{55}] \textbf{Problem:} Answering time depends on answers to poll. That is, \textit{when} the response is sent depends on the real answers. Hence, the answering time represents a \textit{timing channel}.
\end{enumerate}
\begin{enumerate}  
    \item[$\triangleright$] \textbf{Implementation:} Version 2 (\Cref{code:v2})
    \item[$\triangleright$] \textbf{Reasoning:} 
    By using Javascript's \textit{setTimeout(time, function)} we are able to call \textit{submit()} in constant time. Essentially,  \textit{setTimeout} sleeps for a given time and then executes the function. Furthermore, even though the respondent can press a submit button, that button does not trigger a POST. Consequently, even if a respondent finishes the poll before timeout occurs, the client still waits for the timeout.  
    \item[\ding{55}] \textbf{Problem:} If timeout happens before the respondents answers the full poll, the collector learns which questions were not answered. Hence, we need to populate all unanswered questions.
\end{enumerate}
\begin{enumerate}  
    \item[$\triangleright$] \textbf{Implementation:} Version 3 (\Cref{code:v3-short}, full version \Cref{code:v3})
    \item[\ding{55}] \textbf{Problem:} The \textit{amount} of unanswered questions will create a difference in timing. That is, the more unanswered questions, the more times line 3-4 will be executed. Hence, filling in unanswered questions creates a new \textit{timing channel}.  
\end{enumerate}
\begin{enumerate}  
    \item[$\triangleright$] \textbf{Implementation:} Version 4 (\Cref{code:v4-short}, full version \Cref{code:v4})
    \item[$\triangleright$] \textbf{Reasoning:} By pre-populating \textit{each} question with a random answer, we create a loop that executes in constant time (only depends on the amount of questions, which is not secret).
\end{enumerate}

\FloatBarrier

\section{Privacy Evaluation}\label{sec:solution}
In this section, we will first convince the reader that our implementation is in fact differentially private (\Cref{sec:eval-dp}). Next, we will evaluate end-to-end privacy by investigating the two attack surfaces available to the adversary: the communication and the content of the response sent (\Cref{sec:eval-side-channels}).

\subsection{Differential Privacy}\label{sec:eval-dp}
As we use randomized response which is already a well established differentially private algorithm, we will only check that our implementation is sane. First, we check the validity of $t_1, ..., t_n$ and $r_1, ..., r_n$ on line 5 and 4 respectively (\Cref{code:budget}). Next, we give formal privacy guarantees. From \Cref{code:epsilon} it is clear that we find the biggest ratio for any outputs in the same column of the transition matrix. That is, we calculate the biggest possible difference between two data sets as by the definition of differential privacy (\Cref{def:eDP}). Consequently, our implementation is \eptext-differentially private, with an \eptext\ depending on the amount of answers, and $t_1, ..., t_n$.

Through line 6 (\Cref{code:budget}), we enforce the respondents' budget threshold. Since the server is untrusted by the respondent, the client calculates the value of \eptext\ from the poll structure (line 3 \Cref{code:budget}). The client will not allow the respondent to answer any question if the respondent cannot afford the full poll (line 7 \Cref{code:budget}). Since we assume the value of a respondent's budget is public information, we do not leak any additional information by not answering due to insufficient budget.

From \Cref{sec:leakage} it is clear that the implicit flow introduced by follow-up questions is mitigated through flattening each question tree. To clarify, since questions with any amount of follow-up questions and questions with no follow-up question both return \textit{exactly one response}, they are indistinguishable to the attacker. 
\vspace{-1.75em}
\begin{table}[htb]
    \centering
    \begin{tabularx}{\linewidth}{X|l}
        \textbf{Property} & \textbf{Implementation}  \\\toprule
        Validity of poll & Checked on trusted device \\\hline
        Calculation of \eptext\ & By \Cref{thm:DPRR} on trusted device \\\hline
        Enforcement of budget & Before poll \\\hline
        Follow-up question triggered or untriggered & Indistinguishable \\\bottomrule
    \end{tabularx}
    \caption{Evaluated properties}
    \label{tab:dp_eval}
\end{table}
\vspace{-4em}
\subsection{Side-Channels}\label{sec:eval-side-channels}
Based on our threat model (\Cref{sec:model}), the passive adversary can observe and read any network traffic between a data analyst (or an adversary) and a respondent. Since we already explore the implementation of differential privacy, we now assume that all responses sent have \textit{plausible deniability}. Hence, the adversary cannot learn anything about the true answer from observing a response.

In order to learn the true answers, the adversary hopes to observe differences in communication or responses and be able to reverse engineer the true answers. Since we are trying to prove the absence of side-channels, our goal is to exhaustively show all possible cases where true answers could cause differences in communication and poll answers, and refute the possibility of them arising in \Randori. Thus, our objective is to make sure different answers are \textit{indistinguishable} to the adversary.

\begin{figure}[htb]
     \centering
     \begin{subfigure}[b]{0.3\textwidth}
        \centering
        \includegraphics[width=\linewidth]{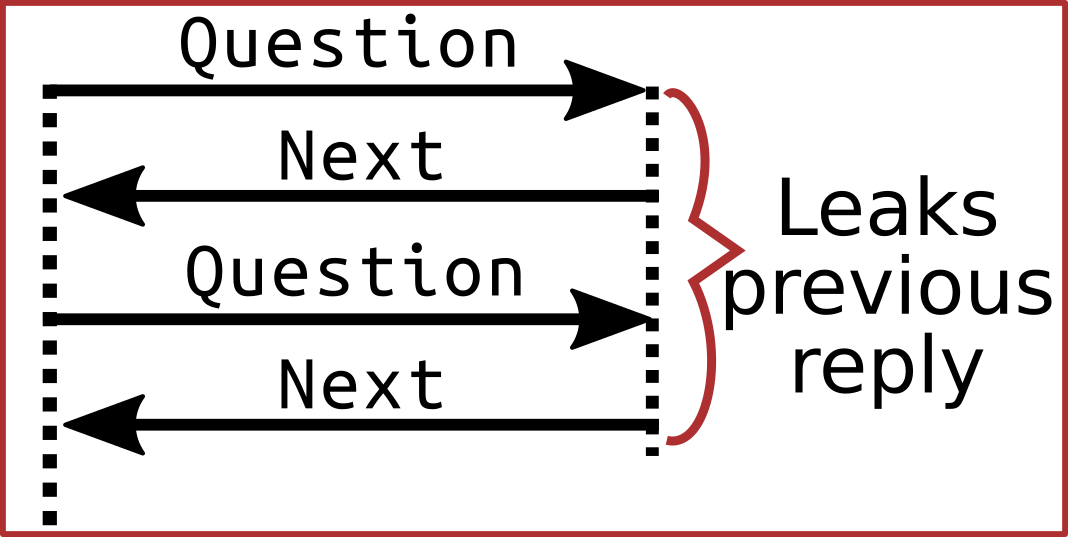}
        \caption{The adversary can learn true answers to questions if a respondent requests (or does not request) follow-up questions}
        \label{fig:communication}
     \end{subfigure}
     \hfill
     \begin{subfigure}[b]{0.3\textwidth}
        \centering
        \includegraphics[width=\linewidth]{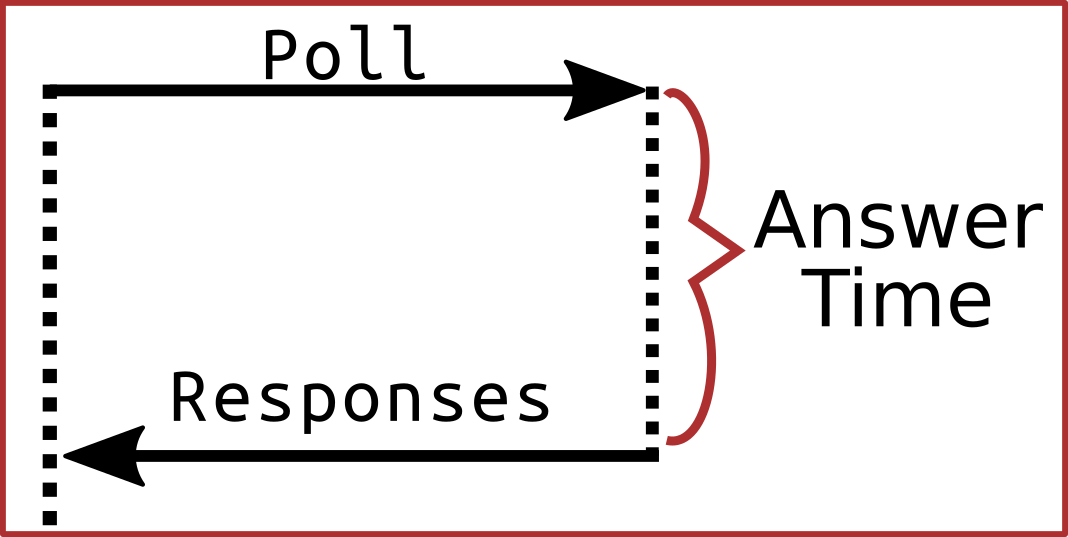}
        \caption{From observing when a poll is received and when the responses are sent, the adversary learns the answering time}
        \label{fig:timing}
     \end{subfigure}
     \hfill
     \begin{subfigure}[b]{0.25\textwidth}
        \centering
        \begin{lstlisting}[language=json, numbers=none, label=fig:unanswered, caption= Reply JSON,basicstyle=\fontsize{8}{9}\selectfont]
"Q1" : "Answer"
"F1" : "Answer"
...
"Fn" : "Answer"
"Q2" : 
\end{lstlisting}
\caption{Illustration of a poll response with unanswered questions}
     \end{subfigure}
     \caption{The three identified cases}
     \label{fig:cases}
\end{figure}

There are two attack surfaces available to the adversary: the communication itself and the message content. We have identified three cases (\Cref{fig:cases}), which we walk through next.
\vspace{-1.75em}

\subsubsection{Case A}\label{sec:communication}
\begin{enumerate}  
    \item[$\triangleright$] \textbf{Attack surface: } communication. 
    \item[$\triangleright$] \textbf{Adversary goal:} learn which follow-up questions the respondent triggers, to deduce answer to previous question.
    \item[$\triangleright$] \textbf{Example attack:} irrelevant and \textit{different} follow-up questions to each answer. That is, the adversary would be able to observe which questions the respondent requests from the server (\Cref{fig:communication}).
\end{enumerate}

There are only two types of communication between the \respondentui\ and the \resultui\ in our implementation: 1) poll request and 2) response post. We need to make sure that the number of GET and POST messages are not related to the respondent's true answers.

\textbf{Mitigation:} Always send the full poll. Our implementation does not allow the respondent's client to send any data when requesting a poll (\Cref{code:v4} line 5) thus requesting anything but the whole poll is impossible. Also, the \respondentui\ only replies with one POST containing all responses at once. Hence, the scenarios next listed are indistinguishable by design:
\vspace{-1.em}
\begin{enumerate}  
    \item[$\circ$] \textit{Respondent requests all questions}
    \item[$\circ$] \textit{Respondent requests some questions}
    \item[$\circ$] \textit{Respondent requests no questions}
\end{enumerate}
\vspace{-1.75em}
\subsubsection{Case B}\label{sec:timing}
\begin{enumerate}  
    \item[$\triangleright$] \textbf{Attack surface: } communication.
    \item[$\triangleright$] \textbf{Adversary goal:} learn one specific answer.
    \item[$\triangleright$] \textbf{Example attack:} many follow-ups for \textit{one} specific answer. That is, the adversary will be able to observe differences in how long time it takes for the respondent to finish the poll (\Cref{fig:timing}). Here, longer answering time means the follow-up was triggered.
\end{enumerate}

There could be different reasons for differences in answering time, and while it may not be possible for the attacker to say with 100\% certainty that the answering time is because of the adversary's follow-ups being triggered, the adversary will be able to shift the probability of this being true. Thus, the adversary would be able to gain an advantage.

Consequently, we want to make any reason for differences in timing indistinguishable to an attacker, such that differences in timing do not leak any additional information.

\textbf{Mitigation:} timeout assures constant answering time, since \textit{submit()} is triggered by the timeout (\Cref{code:v4}, line 9). Furthermore, the same amount of instructions are executed (\Cref{code:v4}, line 16-17 vs line 19-20) whether the question has been answered or a random pre-populated answer is used. What's more, the for-loop is over \textit{var random}, which is of constant size as it contains all question in the poll. Lastly, since the adversary cannot examine the respondent's hardware, they cannot distinguish between the paths in the if-else. Next, we list the differences in timing our implementation takes into account and mitigates:
\begin{enumerate}  
    \item[$\circ$] \textit{Respondent triggers no follow-ups}
    \item[$\circ$] \textit{Respondent triggers some follow-ups}
    \item[$\circ$] \textit{Respondent triggers all follow-ups}
    \item[$\circ$] \textit{Respondent answers fast, not related to follow-up}
    \item[$\circ$] \textit{Respondent answers slowly, not related to follow-ups}
\end{enumerate}
\vspace{-1.75em}
\subsubsection{Case C}\label{sec:message-content}
\begin{enumerate} 
    \item[$\triangleright$] \textbf{Attack surface:} message content. 
    \item[$\triangleright$] \textbf{Adversary goal:} learn one specific answer.
    \item[$\triangleright$] \textbf{Example attack:} many follow-ups for \textit{one} specific answer which cause the respondent to timeout before answering the last question (\Cref{fig:unanswered}). No answer to the last question means the follow-ups were triggered. Note that this specific attack is introduced by our need to use a timeout.
\end{enumerate}
% ...

Since the request for the poll contains no data entered by the respondent, the only place for possible information leakage is through the response post. As each response to a question benefits from plausible deniability due to randomized response, the actual response leak no information. However, unanswered questions would indeed leak if the respondent answered the question or not. Accordingly, the adversary could learn something by observing how many and which questions are unanswered/answered in the response message. 

\textbf{Mitigation:} Since our implementation ensures that each question will have exactly one answer (\Cref{code:v4} line 14-27), the adversary cannot learn anything new from observing which questions are answered/unanswered. Next, we iterate through all different scenarios where the amount of answered questions could differ:
\vspace{-1.em}
\begin{enumerate}  
    \item[$\circ$] \textit{Respondent answers no questions}
    \item[$\circ$] \textit{Respondent answers some questions}
    \item[$\circ$] \textit{Respondent answers all questions}
\end{enumerate}
\vspace{-1.75em}

\section{Discussion, Limitations and Future Work}\label{sec:discussion}
We believe one of the key issues for data analysts to start using differential privacy is understanding the accuracy loss differential privacy causes. In our setting, there is a potential for \textit{dependence between answers} due to the fact that we allow follow-up questions which makes reasoning about accuracy more complicated. For example, recalling the question \myquestion, answering anything to the follow-up question implies that you were unhappy with your purchase. Hence, each follow-up answer adds accuracy to their parent answer. Then, since we allow the data analysts to create polls with arbitrary amount of follow-ups, each answer can have complex dependencies. As such, \textit{it is no longer clear that uniform randomized response is optimal} (i.e. gives each answer equal accuracy), since accuracy is skewed by the fact that the relationships between answers is inherently known by design. 

Consequently, we allow for \textit{non-uniform diagonals} in our transition matrix. While this gives the data analyst more freedom to properly weight their accuracy among answers, it also makes understanding the error more difficult. Hence, we show a Chernoff bound per answer, but this also means that the parameters ($\alpha, \beta, n$) also needs to be tweaked per answer. So while we let the data analyst explore the estimated error, we also believe that analytical error bounds may be too blunt for complex polls. Thus, extending \Randori\ to include empirical error evaluation remains an open and interesting challenge. In fact, we are currently working on a \textit{simulation environment} that allows for this kind of empirical evaluation.

As for trust, we acknowledge that the respondent's receive their client code from the untrusted server. Since the source code of the client is released open source, we assume that the respondent would trust a third party to verify the client code. However, we do not perform any third party checks before letting the respondent answer the poll at this time. A possible and simple extension would be to let a third party serve the client code, and the data analyst would just send the poll.

Regarding the respondent not having a persistent application: this raises two problems. First of all, we do not have a way to save budgets between sessions. We have noted this in our system limitations, but in a real setting this of course becomes a problem, especially when dealing with multiple polls. Our intention is for \Randori's \respondentui\ to be part of an already existing system, for example a web page where the respondent already has an account, which is why we left persistent budgets out of scope. Still, it is important to remember that the respondent's budget needs to be held and updated by a system that the respondent trusts.

Secondly, since the respondent does not have a persistent application, the timeout fails if the respondent closes their client before timeout happens. When the timeout fails, the analyst will not get any response, and as such the analyst's answer may become less accurate than expected (since the prediction is based on $n$ answers, not $n-1$). With a persistent application, the timeout could be paused and resumed if the respondent closes the application. Hence, the data analysts could possibly lose less answers, given that the respondents opened their application again. While this new behavior would change our setting, it would only allow adversaries to learn when the respondent used the application, and not anything about the respondent's true answers. Consequently, allowing the timeout to be kept alive across sections could be beneficial to the data analyst in terms of accuracy, but does not change the respondent's privacy guarantees. However, implementing a persistent timeout process that the respondent could not (accidentally nor maliciously) kill would allow us to loosen our assumptions by no longer requiring ``the respondent does not close the poll before timeout occurs''.

We also acknowledge that the timeout the client uses to avoid timing side-channels introduces a new problem area: if it is too long, we risk that the participant closes the client too early, and if it is too short, the participant might not have time to answer all questions. We do not provide a definitive answer as to what is the best value for this timeout. The problem is mainly that deciding on an optimal value for a timeout is case dependent, and thus very difficult to give a general answer to. 

Lastly, one entirely unexplored area of \Randori\ is usability. So far, we present \Randori\ as \textit{a way} of making differential privacy more accessible to data collectors, as opposed to \textit{the optimal way}. We have also chosen to focus on making differential privacy usable for the data analysts. Hence, interesting next steps for future work include \textit{user studies} 1) where \textit{real data analysts} collect data using \Randori, and 2) where we collect data from \textit{real respondents}. In particular, it would be interesting to let the respondents control their own privacy budget. That is, which values of \eptext\ they are comfortable with before they start answering the polls. As of now, the client only calculates the \eptext, but does not enforce a `useful' (customized) value of \eptext\ in relation to the respondent.
\section{Related Work}\label{sec:related-work} 

The real world example that is most similar to \Randori\ based on what data is collected is the US Census Bureau's deployment of differential privacy \cite{garfinkel_report_nodate}. Even though we collect similarly structured data, a big difference is that the Census Bureau's implementation has been tailored to specific data and therefore deploys release mechanisms under centralized differential privacy.

Several other applications have achieved end-to-end privacy by using local differential privacy, for example applications by Google~\cite{erlingsson_rappor_2014}, Apple~\cite{thakurta_learning_2017-1,thakurta_emoji_2017,differential_privacy_team_apple_learning_2017} and Microsoft~\cite{ding_collecting_2017}. A key difference between \Randori\ and these application is \textit{how} we choose to gather data. Hence, interacting with respondents and gathering inherently dependent data makes \Randori\ novel in comparison.

Next up, work that focuses on making differential privacy accessible. First, the Haskell library \dpella~\cite{lobo-vesga_programming_2020} is similar to \Randori\ when it comes to our use of Chernoff bounds for exploring accuracy. Still, \dpella\ is intended to be used by programmers, and assumes that the data is already stored in a database. \dpella\ is also not limited to randomized response as opposed to \Randori.

Secondly \ektelo\ shares a similar objective with \Randori\ as far as providing accurate, differentially private algorithms to users. Noted, \ektelo\ is much more general than \Randori, and allows users to define new algorithms. What's more, \ektelo\ also concerns the performance of algorithms, which is something we have left completely out of scope in this paper.

Next, \pythia~\cite{kotsogiannis_pythia_2017} has a similar objective as far as providing a tool for \nonexpert s to be able to use differential privacy. \pythia\ is a meta-algorithm that privately finds the most accurate differentially private algorithm to be used for certain input data. Where \pythia\ aims to find the most accurate algorithm, with \Randori\ we aim to help the analyst construct the poll that can give them the most accurate results.

Along the same line of reasoning, the entire Harvard privacy tools project~\cite{harvard_university_privacy_tools_project_harvard_nodate} shares a common goal with \Randori\ as far as making privacy accessible. The project's currently released software focusing on differential privacy are \opendp\ and \rwpsi. \opendp\ is a combination of open tools and end-to-end differentially private systems. As such, \Randori\ could have very well been a part of \opendp\ as \Randori\ is both open source and an end-to-end differentially private system. \rwpsi\ on the other hand is a system for sharing collected data under differential privacy. In essence, \Randori\ and \rwpsi\ only share accessibility as a goal, as \Randori\ is focused on private data collection whereas \rwpsi\ is intended for centralized data sharing.

Other differentially private systems include \airavat~\cite{roy_airavat_2010} and \gupt~\cite{mohan_gupt_2012}. \airavat\ is a distributed system catered for non-privacy experts. Since \airavat\ is a complete system, the authors also focuses on several other issues, such as performance and access control. In a similar manner, \gupt\ is a system intended for enforcing centralized differential privacy on large volumes of data, to be used by \nonexpert s. Our work is only similar through the shared goal of making differential privacy more accessible.

Lastly, we note that there exist several programming languages that make differential privacy accessible to programmers. Among these are \pinq~\cite{mcsherry_privacy_2009}, \fuzz~\cite{reed_distance_2010} and the framework \pretpost~\cite{ebadi_dynamic_2018,ebadi_pretpost_2018}, all intended to be used by experienced programmers. In comparison, our work is more general from an accessibility perspective, by focusing on aiding \nonexpert s that are not necessarily programmers to use differential privacy. 
\section{Conclusion}\label{sec:conclusion}
We implement \Randori, a set of \tools\ for poll design and data collection under differential privacy. A novel part of \Randori\ is that we include the data collection process when reasoning about privacy, and hence we also protect against implicit information leakage. What's more, we make \Randori\ available for all by releasing it as open source software, in order to motivate uninitiated parties to collect data under differential privacy.

To convince the reader that \Randori\ is indeed both differentially private and end-to-end private, we show that our implementation adheres to differential privacy through investigating the code line by line. Then, we evaluate and address how we protect polls from implicit information flows. Next, we evaluate end-to-end privacy by systematically walking through each attack surface and eliminate potential attacks. Consequently, through \Randori, we have made three contributions that map to our originally identified problems. Namely, we provide: 
\begin{itemize}
    \item[\textbf{+}] \tools\ for \textit{designing polls} and \textit{collecting data} under differential privacy
    \item[\textbf{+}] a \tool\ for \textit{predicting and tuning accuracy} of a given poll
    \item[\textbf{+}] an \textit{end-to-end private} implementation of randomized response in a server-client setting
\end{itemize}
\vspace{-1.5em}
%
%
\bibliographystyle{splncs04}
\bibliography{references}

\newpage
\appendix
\section*{Appendix}

\subsection*{Pseudo Code}

\minipage{\linewidth}
\begin{lstlisting}[language=javascript, numbers=left,xleftmargin=2em,frame=single,framexleftmargin=1.75em, label=code:rr, caption=Randomized response as pseudo-code,basicstyle=\fontsize{8}{9}\selectfont]
var transitions = pollToMatrix();
function rr(answers){
 for(answer in answers){
  // Find output space
  let outputs = {};
  //Use transitions to get
  // probability per output
  //ranges[include, exclude]->output
  let ranges = {};
  //Use cryptographic random [1,gcd]
  let random = getRandomInt(1,gcd);
  outputs[answer] = ranges[random];
 }
}
\end{lstlisting}
\endminipage

\minipage{\linewidth}
\begin{lstlisting}[language=javascript, numbers=left,xleftmargin=2em,frame=single,framexleftmargin=1.75em, label=code:epsilon, caption=Calculation of \eptext\ as pseudo-code,basicstyle=\fontsize{8}{9}\selectfont]
var epsilon = undefined;
// For each question subtree
  let potential_epsilon = undefined;
  // For each answer
  let max = undefined;
  let min = undefined;
    // Loop through all other answers
    // Get max probability ratio
    let check=Math.max(max.div(min),
        min.div(max));
    // Bigger?
    if(potential_epsilon==undefined 
      || potential_epsilon<check){
      potential_epsilon = check;
    }
  epsilon+=potential_epsilon;
\end{lstlisting}
\endminipage

\minipage{\linewidth}
\begin{lstlisting}[language=javascript, numbers=left,xleftmargin=2em,frame=single,framexleftmargin=1.75em, label=code:budget, caption=Enforcement of budget thershold,basicstyle=\fontsize{8}{9}\selectfont]
var budget = 100; // ln(budget)
var truth_threshold = 0.99;
var cost = calculateEpsilon();
var ok_truth = withinThreshold();
if(cost > budget){
  //Disable UI
} else if(!ok_truth){
  //Disable UI
}else {
  budget-=cost;
  // Show poll
}
\end{lstlisting}
\endminipage

\minipage{\linewidth}
\begin{lstlisting}[language=javascript, numbers=left,xleftmargin=2em,frame=single,framexleftmargin=1.75em,label=code:v3,caption=Version 3,basicstyle=\fontsize{8}{9}\selectfont]
var answers = {};
var timeout = 9000; //Example
fetch('/poll');
setTimeout(submit, timeout);
...
// Respondent answers poll
...
function submit(){
    for(answer in answers){
        if(answers[answer]==null){
            answers[answer]=random();
        }
    }
    let responses = rr(answers);
    fetch('/submit', {
      method: 'POST',
      body: JSON.stringify(responses)
      });
}
\end{lstlisting}
\endminipage

\minipage{\linewidth}
\begin{lstlisting}[language=javascript, numbers=left,xleftmargin=2em,frame=single,framexleftmargin=1.75em, label=code:v4,caption=Version 4,basicstyle=\fontsize{8}{9}\selectfont]
var answers = {};
var shadow = {};
var random = {};
var timeout = 9000; //Example
fetch('/poll');
for(question in poll){
    random[question]=random();
}
setTimeout(submit, timeout);
...
// Respondent answers poll
...
function submit(){
    for(answer in random){
        if(shadow[answer]==null){
            answers[answer]=
            random[answer];
        } else {
            answers[answer]=
            shadow[answer];
        }
    }
    let responses = rr(answers);
    fetch('/submit', {
      method: 'POST',
      body: JSON.stringify(responses)
      });
}
\end{lstlisting}
\endminipage

\newpage
\subsection*{Graphical Interfaces}

\begin{figure}[htb]
    \centering
    \includegraphics[scale=0.7]{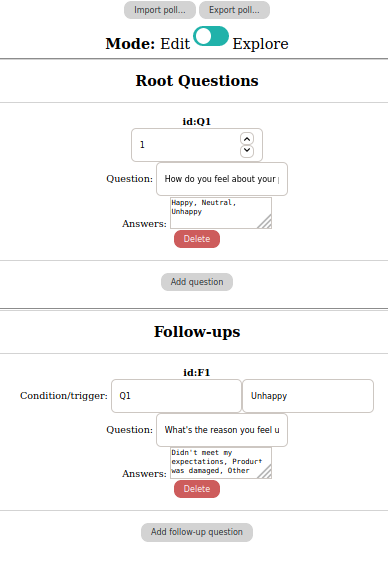}
    \caption{The edit mode of the \editor}
    \label{fig:edit}
\end{figure}

\begin{figure}[htb]
    \centering
\begin{lstlisting}[language=json, label=fig:json, caption=JSON format of the example question,basicstyle=\fontsize{8}{9}\selectfont]
{"children":[
  {"qid":"F1",
  "question":"What's the reason you feel unhappy?",
  "answers":["Didn't meet my expectations"," Product was damaged"," Other"],
  "probability":["1/3","1/3","1/3"]}
],
"roots":[
  {"qid":"Q1",
  "truth":"1/2",
  "question":"How do you feel about your purchase?",
  "answers":["Happy","Neutral","Unhappy"],
  "probability":["1/3","1/3","1/3"]}
],
"paths":[["Q1","Unhappy","F1"]],
"order":["Q1"]
}
\end{lstlisting}
    \caption{The JSON produced by the example question}
    \label{fig:example_json}
\end{figure}

\begin{figure}[htb]
     \centering
     \includegraphics[width=\linewidth]{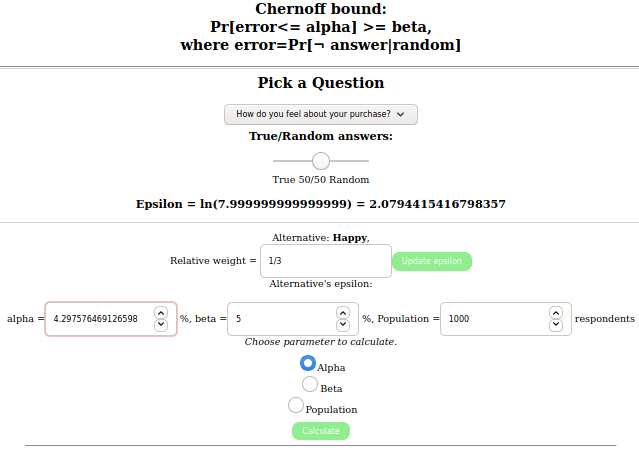}
        \caption{The explore mode of the \editor}
        \label{fig:explore}
\end{figure}
\begin{figure}[htb]
         \centering
         \includegraphics[scale=0.5]{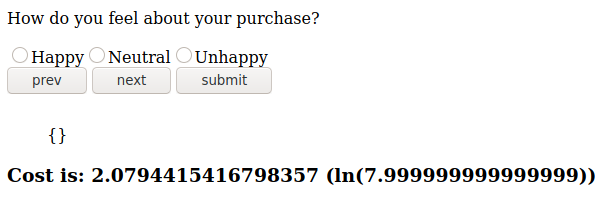}
        \caption{The respondent's view of the poll, here explicitly showing \eptext\ and chosen answers. Note that fractions are used in calculations internally, but here we are showing cost as floating point numbers.}
        \label{fig:respondent-ui}
\end{figure}

\end{document}